\documentclass[10pt,twocolumn]{IEEEtran}
\usepackage{graphicx}
\usepackage{amsmath}
\usepackage{amssymb}
\usepackage[caption=false]{subfig}
\usepackage[noadjust]{cite}
\usepackage{float}
\usepackage{color}
\usepackage{url}

\newtheorem{theorem}{Theorem}
\begin{document}

\bstctlcite{IEEEexample:BSTcontrol}

\title{Non-Orthogonal Multiple Access for Mobile VLC Networks with Random Receiver Orientation}

\author{Yavuz Yap{\i}c{\i} and \.{I}smail G\"{u}ven\c{c} \\
Department of Electrical and Computer Engineering, North Carolina State University, Raleigh, NC\\
{\tt \{yyapici,iguvenc\}@ncsu.edu}%
\thanks{This work is supported in part by NSF CNS award 1422062.}}

\maketitle

\begin{abstract}
We consider a downlink multiuser visible light communications (VLC) network where users randomly change their location and vertical orientation. The non-orthogonal multiple access (NOMA) strategy is adopted to serve multiple users simultaneously, and, hence, to improve spectral efficiency. We propose two novel user scheduling schemes for NOMA, which are referred to as \textit{individual} and \textit{group-based}. In order to further reduce the computational complexity and link overhead, novel \textit{limited-feedback} schemes (on channel quality) are also proposed, which basically involve mean vertical angle (instead of its instantaneous value). Moreover, a two-bit feedback scheme is proposed for group-based user scheduling, which relies on not only distance but also vertical angle (in contrast to conventional one-bit feedback with distance only). The outage probability and sum-rate expressions are derived analytically, which show a very good match with the simulation data. Numerical results verify that the practical feedback scheme with the mean vertical angle achieves a near-optimal sum-rate performance, and the two-bit feedback significantly outperforms the one-bit feedback.
\end{abstract}


\section{Introduction}\label{sec:intro}

Visible light communications (VLC) is a promising technology for wireless 5G networks and beyond by leveraging the broad license-free optical spectrum at wavelengths of $380$-$750$ nm~\cite{Richardson13VLC}. Together with developments on light emitting diode (LED) as the primary illumination source, VLC networks appear as a viable solution for simultaneous illumination and communication at low power consumption and with high durability~\cite{Yapici2017MulEleVLC,Eroglu2018SofDef,Haas14VLCBeyond}. The non-orthogonal multiple access (NOMA) appears as a powerful technology for multiuser VLC networks, as well, which suggests to serve multiple users at the same time and frequency slot~\cite{Ding2017AppNoma, Poor2017NomaMul, Dobre2017PowDomNoma}.

The NOMA strategy has been considered for VLC networks with a limited attention. In~\cite{Uysal2015Noma}, NOMA is considered in a VLC scenario where the respective performance is compared to that of the orthogonal frequency division multiple access (OFDMA) scheme. The performance analysis of NOMA is performed in \cite{Haas2016PerEvaNoma, Karagiannidis2016Noma} for VLC networks along with lighting quality and power allocation. For a VLC NOMA system, a multiple-input multiple-output (MIMO) setting is explored in~\cite{Du2017OnPeMimo}, bit-error-rate (BER) analysis is performed in~\cite{Karagiannidis2017OnPerNoma}, sum-rate maximization is conducted in~\cite{Li2017FaiNoma}, a location based user grouping scheme is offered in~\cite{Xu2017UseGro}, and a phase pre-distorted  symbol detection method is proposed in~\cite{Cha2017JoiDet}.

As a major drawback, VLC transmission highly relies on LOS links, which may not be readily available all the time. This problem is pronounced even more for dynamic VLC scenarios involving random receiver orientations~\cite{Haas2016AccPoiSel, Haas2017HanModInd, Yapici2017VerOriC, Yapici2017RanVerOr, Chen2017ImpBer, Wang2017ImpRec, Haas2018ImpTerOri, Haas2018ModRanOri}. In \cite{Haas2016AccPoiSel}, a new metric for the access point selection problem is proposed for receivers with random orientations. In \cite{Yapici2017VerOriC}, a general framework for random receiver orientation is developed where the square-channel gain distribution is derived analytically. The proposed framework applies to any prior distribution for the receiver orientation, and the approach is generalized to multi-LED scenarios in \cite{Yapici2017RanVerOr}. The impact of tilting the receiver angle on the BER performance is considered in \cite{Chen2017ImpBer}, and further studied in \cite{Wang2017ImpRec} to yield capacity bounds. Finally, the distribution of the receiver orientation in light-fidelity (LiFi) downlink networks is evaluated in \cite{Haas2018ModRanOri} through indoor measurements. 

In this paper, we consider a multiuser VLC network where mobile users having random location and vertical orientation are served simultaneously by novel limited-feedback NOMA strategies. To the best of our knowledge, this realistic VLC NOMA scenario has not been studied in the literature before. In particular, we propose two novel NOMA strategies called \textit{individual} and \textit{group-based} user scheduling, which are basically designed to reduce the complexity and feedback overhead. In addition, we also propose to use the \textit{mean} vertical angle (instead of its instantaneous value) as a limited yet sufficient feedback scheme. Moreover, a novel \textit{two-bit feedback} scheme is also proposed as a practical feedback mechanism, which employs both the distance and vertical angle information in one bit each, and differs from the conventional \textit{one-bit} feedback involving distance only~\cite{Ding2017RanBea}. We also derive the analytical expressions for the outage probability and sum rates for each of these NOMA strategies, which show a very good match with the respective simulation data. 

The rest of the paper is organized as follows. Section~\ref{sec:system} introduces the system model. Section~\ref{sec:noma} considers various NOMA strategies in VLC networks, where the respective outage analysis is given in Section~\ref{sec:outage_analysis}. The numerical results are presented in Section~\ref{sec:results}, and the paper concludes with some final remarks in Section~\ref{sec:conclusion}.

\section{System Model} \label{sec:system} 
\begin{figure}[!t]
\centering
\includegraphics[width=0.47\textwidth]{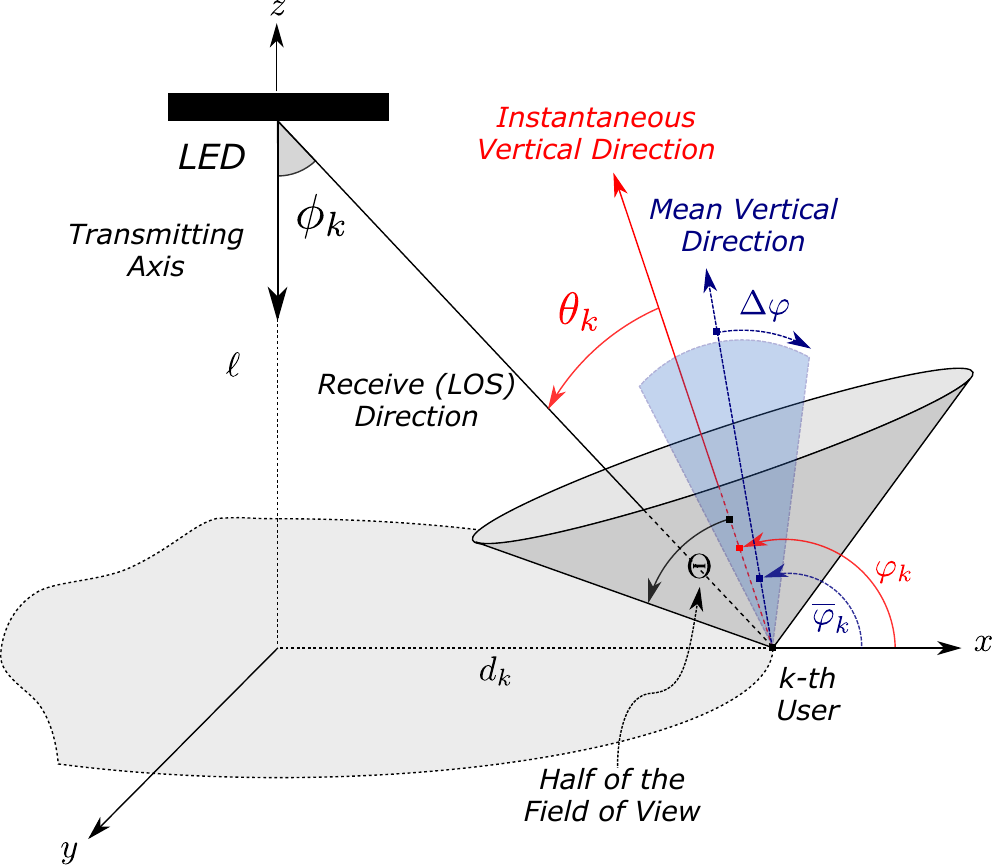}\vspace{-0.0in}
\caption{Multiuser VLC network with $k$th user participating into NOMA.}
\label{fig:setting}\vspace{-0.0in}
\end{figure}

We consider an indoor VLC downlink transmission scenario involving a single transmitting LED and $K$ users. The interaction between the LED and the $k$th user over a LOS link is depicted in Fig.~\ref{fig:setting}, and respective optical direct current (DC) channel gain is represented as~\cite{Haas2018PhySec}
\begin{align}\label{eqn:channel}
h_k =\frac{(m+1)A_r}{2\pi (\ell^2 + d_k^2)}\cos^m( \phi_k ) \cos( \theta_k ) \, \Pi\big[ \left| \theta_k \right| /\Theta \big],
\end{align}
where $\ell$ is the vertical distance between the LED and the plane including all the users, $d_k$ is the horizontal distance of the $k$th user to the LED, $\phi_k$ and $\theta_k$ are the corresponding angle of irradiance and incidence, respectively. The Lambertian order is $m\,{=}\,{-}1/\log_2(\cos(\Phi))$ with $\Phi$ being the half-power beamwidth of the LED, and $A_r$ and $\Theta$ are the detection area and half of the FOV for the photodetectors, respectively. The function $\Pi[x]$ takes $1$ whenever $|x| \leq 1$, and is $0$ otherwise.   

We assume that the users are non-static within both the horizontal and vertical planes such that they are continuously changing their horizontal locations and vertical orientations. In particular, $d_k$ is assumed to follow a Uniform distribution with $\mathcal{U}\,[d_{\rm min}{,}\,d_{\rm max}]$ and $\Delta d \,{=}\,d_{\rm max}{-}d_{\rm min}$. In addition, $\varphi_k$ is also varying around a \textit{mean vertical angle} $\overline{\varphi}_{k}$, which is picked up from a Uniform distribution with $\mathcal{U}\,[\overline{\varphi}_\text{min}{,}\,\overline{\varphi}_\text{max}]$ and $\Delta\overline{\varphi}\,{=}\,\overline{\varphi}_\text{max}{-}\overline{\varphi}_\text{min}$~\cite{Haas2018ModRanOri}. As a result, the $k$th user's orientation or, equivalently, the \textit{vertical angle} $\varphi_{k}$ takes a value from a uniform distribution with $\mathcal{U}\,[\overline{\varphi}_{k}{-}\Delta\varphi{,}\,\overline{\varphi}_{k}{+}\Delta\varphi]$ for a given value of $\overline{\varphi}_k$ and a maximum deviation angle of $\Delta\varphi$. 

Considering Fig.~\ref{fig:setting}, the incidence angle $\theta_k$ is given as
\begin{align}\label{eqn:incidenceang}
\theta_k = \pi - \tan^{{-}1}(\ell/d_k) - \varphi_k ,
\end{align}
where $d_k$ and $\varphi_k$ take values independently (in contrast to \cite{Yapici2017RanVerOr} where they are coupled through $\theta_k$). Note that $\theta_k$ in \eqref{eqn:incidenceang} is allowed to take either positive or negative values depending on the values of $d_k$ and $\varphi_k$. This definition enables a more realistic scenario, where the vertical orientation can take any value regardless of how far the $k$th user is away from the LED. Hence, the incidence angle $\theta_k$ possesses \textit{independent} contributions of $d_k$ and $\varphi_k$.

Furthermore, we assume that $d_k$ and $\overline{\varphi}_{k}$ are varying much slowly as compared to $\varphi_{k}$, and, hence, have relatively \textit{large coherence time}, which well aligns with realistic scenarios~\cite{Haas2018ModRanOri}. In other words, each user is changing its location and mean vertical direction slowly, whereas relatively small variations happen much quickly in actual vertical direction. As a result, we consider $d_k$ and $\overline{\varphi}_{k}$ as good candidates representing the channel state information (CSI), of the $k$th user. Since $d_k$ and $\overline{\varphi}_{k}$ can both be tracked with less computational burden (as compared to $\varphi_k$), we employ these parameters in \textit{limited-feedback schemes} for the NOMA transmission. 

On the other hand, any limited-feedback scheme involving $d_k$ and $\overline{\varphi}_{k}$ (instead of $\varphi_k$) is likely to degrade the user rates as compared to the full CSI feedback. This is, in part, because the combination of $d_k$ and $\overline{\varphi}_{k}$ is not capable of capturing the true status of the receive direction (i.e., being inside or outside the FOV), which we call \textit{FOV status} and represented by $\Pi\left[ |\theta_k|/\Theta \right]$ in \eqref{eqn:channel}. In the subsequent sections, we consider this compromise between the limited-feedback (with lower computational burden and overhead) and the full CSI (with better user rates) mechanisms for the NOMA transmission.

\vspace{-0.15in}
\section{NOMA in VLC Downlink Channels} \label{sec:noma}
In this section, we consider NOMA for a VLC downlink scenario with full CSI and limited-feedback schemes.

\vspace{-0.15in}
\subsection{Sum Rates for VLC NOMA Transmission}\label{sec:noma_userrates}

We assume that the users are ordered in ascending order such that $i$th user has the $i$th smallest nonzero channel gain. Moreover, we assume that $L$ users are involved in NOMA transmission out of $K$ users with $L\,{\leq}\,K$, and that $\mathcal{S}$ is the set including indices of these $L$ NOMA users. The signal-to-interference-plus-noise ratio (SINR) at the $j$th user while decoding the message of a weaker $i$th user with $i\,{<}\,j\,{\leq}\,K$ is 
\begin{align}\label{eqn:sinr_ij}
\mathsf{SINR}_{i{\rightarrow}j} = \frac{h_j^2 \beta_i^2}{h_j^2 \sum\limits_{k\in\mathcal{S}_i}\beta_k^2 + \gamma^{{-}1}},
\end{align}
where $\mathcal{S}_i$ is the set involving indices of the users being stronger than the $i$th user, $\beta_k$ is the optical power allocation coefficient of the $k$th user such that $\sum_{k{\in}\mathcal{S}}\beta_k^2\,{=}\,1$, and $\gamma$ is the equivalent electrical transmit signal-to-noise ratio (SNR). Note that \eqref{eqn:sinr_ij} implicitly assumes that the message of any $k$th user with $k\,{<}\,i$ (i.e., having a relatively weaker channel gain) has already been decoded successfully, and subtracted from the received signal as per successive interference cancellation (SIC) approach. 

The SINR of the $j$th user while decoding its message is
\begin{align}\label{eqn:sinr_j}
\mathsf{SINR}_{j} = \frac{h_j^2 \beta_j^2}{h_j^2 \sum\limits_{k\in\mathcal{S}_j}\beta_k^2 + \gamma^{{-}1}},
\end{align}
where $\mathsf{SINR}_{\kappa}\,{=}\,h_\kappa^2 \beta_\kappa^2 \gamma$ for $\kappa$ being the index of the strongest NOMA user (i.e., $\mathcal{S}_\kappa$ is an empty set). At any NOMA user, the overall decoding mechanism is assumed to be in outage if instantaneous user rates associated with either of \eqref{eqn:sinr_ij} or \eqref{eqn:sinr_j} do not meet the respective target rates of the NOMA users based on their preset quality-of-service (QoS) requirements. 

Note that the conventional Shannon formulation does not hold for VLC links since the optical signal has certain average and peak power constraints as well as being non-negative. We therefore consider $R_{i{\rightarrow}j}\,{=}\frac{1}{2}\,\log_2 \left( 1 \,{+}\, \frac{e}{2\pi}\mathsf{SINR}_{i{\rightarrow}j} \right)$ and $R_{j}\,{=}\,\frac{1}{2}\log_2 \left( 1 \,{+}\, \frac{e}{2\pi}\mathsf{SINR}_{j} \right)$  as an instantaneous achievable rate pair associated with \eqref{eqn:sinr_ij} and \eqref{eqn:sinr_j}, respectively,~\cite{Haas2018PhySec}. The outage probability of the $j$th user is therefore given as 
\begin{align}\label{eqn:outage_j_1}
\mathsf{P}_{j}^{\rm o}  = 1 - \Pr \Big( \bigcap\limits_{k\in\overline{\mathcal{S}}_j} R_{k{\rightarrow}j} > \overline{R}_k, \, R_{j} > \overline{R}_{j}\Big),
\end{align}
where $\overline{R}_k$ is the QoS based target rate of the $k$th user, and $\overline{\mathcal{S}}_j$ is the set involving indices of the users being weaker than the $j$th user. Defining $\epsilon_k\,{=}\,\left( 2^{2{\overline{R}_k}}\,{-}\,1 \right) { \frac{2\pi}{e}}$, \eqref{eqn:outage_j_1} becomes
\begin{align}\label{eqn:outage_j_2}
\mathsf{P}_{j}^{\rm o}  = 1 - \Pr \Big( \bigcap\limits_{k\in\overline{\mathcal{S}}_j} \mathsf{SINR}_{k{\rightarrow}j} > \epsilon_k, \, \mathsf{SINR}_{j} > \epsilon_{j}\Big),
\end{align}
and, the respective sum-rate expression is
\begin{align}\label{eqn:sumrate_noma}
R^{\rm NOMA} = \displaystyle\sum_{k{=}1}^{L} \left( 1 - \mathsf{P}_{k}^{\rm o} \right) \overline{R}_k .
\end{align}

For the OMA transmission, all resources are allocated to a single user being served during $1/L$ of the transmission period, and hence the sum-rate expression is
\begin{align}\label{eqn:sumrate_oma}
R^{\rm OMA} = \displaystyle\sum_{k{=}1}^{L} \left[ 1 - \Pr \left( \left| h_k \right|^2 \leq \left( 2^{2{\overline{R}_k}}\,{-}\,1 \right) \frac{2\pi}{e} \right) \right] \overline{R}_k .
\end{align}

\vspace{-0.15in}
\subsection{Individual User Scheduling with Limited Feedback}\label{sec:noma_feedback_individual}
In this NOMA strategy, users to be served simultaneously are chosen among a total of $K$ users based on their \textit{individual} channel qualities. It is therefore vital for the NOMA transmitter to order \textit{all} the potential users according to their channel qualities based on the information transmitted back from the users. The optimal strategy from this perspective is therefore to order the users based on their full CSI given as follows
\begin{align}\label{eqn:order_fullcsi}
\left| h_1 \right|^2 \leq \left| h_2 \right|^2 \leq \dots \leq \left| h_K \right|^2.
\end{align}  

It is, however, not practical to employ the order in \eqref{eqn:order_fullcsi} as it necessitates the channel gains to be tracked continuously. Since $d_k$ and the mean vertical angle $\overline{\varphi}_{k}$ is varying slower as compared to \textit{instantaneous} vertical angle $\varphi_k$, we consider a limited-feedback mechanism where $d_k$ and $\overline{\varphi}_{k}$ (not $\varphi_k$) are sent back to the NOMA transmitter. This mechanism circumvents the necessity of continuous tracking of $\varphi_{k}$, and hence relieves the computational complexity. The transmitter employs the following order while choosing NOMA users
\begin{align}\label{eqn:order_limited}
\left| \overline{h}_{\vartheta_1} \right|^2 \leq \left| \overline{h}_{\vartheta_2} \right|^2 \leq \dots \leq \left| \overline{h}_{\vartheta_K} \right|^2,
\end{align} 
where $\overline{h}_{\vartheta_k}$ is the \textit{average} DC channel gain of the user with the index $\vartheta_k$, which has the $k$th smallest average DC channel gain among all, and is given as
\begin{align}\label{eqn:channel_mean}
\overline{h}_k =\frac{(m+1)A_r}{2\pi (\ell^2 + d_k^2)} & \cos^m( \phi_k ) \left| \cos(\tan^{{-}1}(\ell/d_k) + \overline{\varphi}_k ) \right| \, \nonumber \\
& \times \Pi\big[ \left| \pi - \tan^{{-}1}(\ell/d_k) - \overline{\varphi}_k \big| /\Theta \right],
\end{align}
for $k\,{\in}\,\left\lbrace \vartheta_1,\vartheta_2,\dots,\vartheta_K\right\rbrace$. This definition also implies that $\overline{h}_{\vartheta_k}$ does not necessarily appear as the $k$th record in the full CSI feedback based order of \eqref{eqn:order_fullcsi}. Note that this limited-feedback mechanism cannot provide the correct FOV status to the NOMA transmitter (as mean value $\overline{\varphi}_k$ is sent instead of instantaneous value $\varphi_k$), which is likely to end up with sum-rate performance degradation. Note also that this performance loss remains marginal as $\Delta\varphi$ (between $\overline{\varphi}_k$ and the maximum value of $\varphi_k$) gets smaller.    

\subsection{Group-Based User Scheduling with Two-Bit Feedback}\label{sec:noma_feedback_group}
We now consider a different NOMA strategy where the transmitter does not order all $K$ potential users individually, but rather groups them based on \textit{two-bit} information on their channel qualities. The feedback mechanism this time sends \textit{low-rate} information being composed of two bits, which represent either 1) $d_k$ and $\varphi_{k}$, or 2) $d_k$ and $\overline{\varphi}_{k}$. In either case, the distance and (average) incidence angle are compared to their own preset threshold values, and the result is transmitted back to the transmitter in two bits of information.

More specifically, the respective feedback bits are $\left\lbrace \Pi \left( d_k/d_{\rm th} \right),\Pi \left( \left|\theta_k\right|/\theta_{\rm th} \right) \right\rbrace$ if $d_k$ and $\varphi_{k}$ are both available to the user $k$ for feedback computation, where $d_{\rm th}$ and $\theta_{\rm th}$ are threshold values. If the user $k$ has the information of $d_k$ and $\overline{\varphi}_{k}$, then the feedback bits become $\left\lbrace \Pi \left( d_k/d_{\rm th} \right),\Pi \left( \left|\overline{\theta}_k\right|/\theta_{\rm th} \right) \right\rbrace$ with $\overline{\theta}_k \,{=}\, \pi {-} \tan^{{-}1}(\ell/d_k) {-} \overline{\varphi}_k$. 

At the transmitter, all users are split into groups, where each group is composed of users having the same feedback bits. Assuming that $d_k$ and $\varphi_{k}$ are used in feedback computations, the groups of users having weaker and stronger channel gains can be represented, respectively, as follows
\begin{align}
\mathcal{S}_{\rm W, \varphi} &= \left\lbrace k \mid  d_k\,{>}\,d_{\rm th},\,    \left|\theta_k\right| \,{>}\, \theta_{\rm th} \right\rbrace, \label{eqn:set_weak_ins} \\
\mathcal{S}_{\rm S, \varphi} &= \left\lbrace k \mid  d_k\,{\leq}\,d_{\rm th},\, \left|\theta_k\right| \,{\leq}\,\theta_{\rm th} \right\rbrace. \label{eqn:set_strong_ins}
\end{align}
Similarly, whenever $d_k$ and $\overline{\varphi}_{k}$ are used while computing feedback bits, these groups become  
\begin{align}
\mathcal{S}_{\rm W, \overline{\varphi}} &= \left\lbrace k \mid  d_k\,{>}\,d_{\rm th},\, \left|\overline{\theta}_k \right| \,{>}\,\theta_{\rm th} \right\rbrace,  \label{eqn:set_weak_mean} \\
\mathcal{S}_{\rm S, \overline{\varphi}} &= \left\lbrace k \mid  d_k\,{\leq}\,d_{\rm th},\, \left|\overline{\theta}_k \right|  \,{\leq} \,\theta_{\rm th} \right\rbrace. \label{eqn:set_strong_mean}
\end{align}

We confine our search for the NOMA users to the sets $\mathcal{S}_{\rm W, \varphi}$ ($\mathcal{S}_{\rm W, \overline{\varphi}}$) and $\mathcal{S}_{\rm S, \varphi}$ ($\mathcal{S}_{\rm S, \overline{\varphi}}$), which are more likely to involve good candidates for weaker and stronger NOMA users, respectively. We therefore pick up the weaker NOMA user from $\mathcal{S}_{\rm W, \varphi}$ ($\mathcal{S}_{\rm W, \overline{\varphi}}$), and the stronger NOMA user is similarly chosen from $\mathcal{S}_{\rm S, \varphi}$ ($\mathcal{S}_{\rm S, \overline{\varphi}}$). In Section~\ref{sec:results}, we consider one-bit feedback of $d_k$ only~\cite{Poor2017RanBea}, as well, which cannot capture the FOV status at all, and the resulting sum-rate performance is hence much worse.  

\section{Outage Analysis for VLC NOMA}\label{sec:outage_analysis}

In this section, we derive the exact outage probability expressions for the NOMA strategies in Section~\ref{sec:noma}. 
\subsection{Outage Formulation}\label{sec:outage_formulation}
In the VLC downlink transmission, the channel gain in \eqref{eqn:channel} can take either zero or a nonzero value, which depends on the receive direction being inside or outside the FOV. This is a major difference of the VLC transmission from its RF counterparts, and hence we force the NOMA transmitter to schedule only the users having nonzero channel gains. We designate $i$ and $j$ being the index of the users having weaker and stronger \textit{nonzero} channel gains, respectively, with $i\,{<}\,j\,{\leq}\,K$. The outage probability for the $i$th user is given as
\begin{align}
\mathsf{P}_{i}^{\rm o}  &= 1 - \Pr \left( \mathsf{SINR}_{i} > \epsilon_i \mid K_{\rm nz} \geq j \right)  \label{eqn:outpr_i_1} \\
&= 1 - \Pr \left( h_i^2 > \eta_i \mid K_{\rm nz} \geq j \right),  \label{eqn:outpr_i_3}
\end{align}
where $\eta_i\,{=}\,\frac{\epsilon_i/\gamma}{\beta_i^2{-}\beta_j^2\epsilon_i}$, and $K_{\rm nz}$ is the number of users having nonzero channel gain. Similarly, the outage probability for the $j$th user is given as
\begin{align}
\mathsf{P}_{j}^{\rm o}  &= 1 - \Pr \left( \mathsf{SINR}_{i{\rightarrow}j} > \epsilon_i, \, \mathsf{SINR}_{j} > \epsilon_{j} \mid K_{\rm nz} \geq j\right), \label{eqn:outpr_j_1}\\
&= 1 - \Pr \left( h_j^2 > \eta_j \mid K_{\rm nz} \geq j\right),  \label{eqn:outpr_j_3}
\end{align}
where $\eta_j\,{=}\,\max\left\lbrace \frac{\epsilon_i/\gamma}{\beta_i^2{-}\beta_j^2\epsilon_i} ,\, \frac{\epsilon_j/\gamma}{\beta_j^2} \right\rbrace$. Finally, employing \eqref{eqn:outpr_i_3} and \eqref{eqn:outpr_j_3} in \eqref{eqn:sumrate_noma} gives the outage sum rates. Note that for the NOMA strategy in Section~\ref{sec:noma_feedback_group}, we do not have the condition $K_{\rm nz} \,{\geq}\, j$ in \eqref{eqn:outpr_i_3} and \eqref{eqn:outpr_j_3}.

\begin{theorem}\label{lem:numof_nzusers}
$K_{\rm nz}$, number of users having nonzero channel gain, follows Binomial distribution with $\mathcal{B}(K,p)$ where
\begin{align}
p &= \frac{1}{\Delta d} \int_{d_{\rm min}}^{d_{\rm max}} \Delta F_{\varphi} \left(r,\Theta \right) {\rm d}r,\label{eqn:success_prob}
\end{align}
where $\Delta F_{\varphi} \left( x,y \right) \,{=}\, \tilde{F}_{\varphi}\left( x,y \right) \,{-}\, \tilde{F}_{\varphi} \left( x, {-} y  \right)$ is the difference of two CDFs with $\tilde{F}_{\varphi} \left( x,y \right) \,{=}\, F_{\varphi}\left( \pi {-} \tan^{{-}1}(\ell/x) {+} y \right)$ and $F_{\varphi}$ being the CDF of $\varphi$.
The respective probability mass function (PMF) of $K_{\rm nz}$ is given as 
\begin{align}\label{eqn:pmf_knz_mdf}
\hspace{-0.05in}p_{K_{\rm nz}} \left( k | k_{\rm min} \right) = \begin{cases}
\displaystyle c_{\rm nz} \binom{K}{k} p^k (1{-}p)^{K{-}k} & \textrm{if } k \,{\geq}\, k_{\rm min},\\
0 & \textrm{otherwise},
\end{cases}
\end{align}
where $c_{\rm nz} \,{=} \sum_{k=k_{\rm min}}^{K} \!\!\binom{K}{k} p^k (1{-}p)^{K{-}k}$, and $k_{\rm min}$ is the minimum number of users having nonzero channel gain to start the NOMA transmission (i.e., $k_{\rm min}\,{=}\,j$ in the strategy of Section~\ref{sec:noma_feedback_individual}).
\end{theorem}
\begin{IEEEproof}
See~\cite{Yapici2018NonVlc} for a complete proof, which we could not involve herein due to space limitations.
\end{IEEEproof}

\subsection{CDF of Square-Channel for Individual User Scheduling}\label{sec:cdf_individual}
In this section, we present the CDF of the nonzero square-channel gain for individual user-scheduling NOMA transmission. 

\begin{theorem}\label{the:cdf_unordered}
The CDF of the unordered nonzero square-channel is given as
\begin{align} 
F_{h^2|h>0}(x) &= 1 {-} \frac{\displaystyle \int_{d_{\rm min}}^{d_{\rm max}} \Delta F_{\varphi} \left( r,\psi \left(x,r,\Theta\right) \right) {\rm d}r}
{\displaystyle \int_{d_{\rm min}}^{d_{\rm max}} \Delta F_{\varphi} \left( r,\Theta \right) {\rm d}r}, \label{eqn:cdf_unordered}
\end{align}
where $\psi(x,y,z) \,{=}\, \min \left( 1/2\cos^{{-}1}\!\left( 2\min \left(x\upsilon(y),1\right){-}1\right),z \right)$ with $\upsilon(x)\,{=}\,(\ell^2 + x^2)^{m+2}h_c^{{-}2}$ and $h_c^2\,{=}\,(m+1)A_r\ell^m/2\pi$, and $\Delta F_{\varphi} \left( x,y \right)$ is defined in Theorem~\ref{lem:numof_nzusers}.
\end{theorem}
\begin{IEEEproof}
See~\cite{Yapici2018NonVlc} for a complete proof.
\end{IEEEproof}

When we choose the $k$th user having nonzero channel gain, we actually order a total of $K_{nz}$ users since the remaining $K\,{-}\,K_{nz}$ have all zero gain. The CDF of the ordered nonzero square-channel gain of the $k$th user can therefore be found using order statistics~\cite{Nagaraja2005OrdSta} as follows
\begin{align}
F_{h_k^2|h_k>0}(x) &\,{=} \!\!\!\! \sum\limits_{n=k_{\min}}^{K} \sum\limits_{\ell=k}^{n} \binom{n}{\ell} \binom{K}{n} c_{\rm nz} \, p^{n} (1{-}p)^{K{-}n} \nonumber \\
& \hspace{0.2in} \left[ F_{h^2|h>0}(x)\right]^{\ell} \left[ 1-F_{h^2|h>0}(x)\right]^{n-\ell}. \label{eqn:cdf_ordered}
\end{align}
The desired outage probabilities of individual user-scheduling NOMA (i.e., \eqref{eqn:outpr_i_3} and \eqref{eqn:outpr_j_3}) can then be computed using \eqref{eqn:cdf_ordered}.

\subsection{CDF of Square-Channel for Group-Based User Scheduling}\label{sec:cdf_twobit}

We finally present the CDF of the nonzero square-channel gain for two-bit feedback NOMA with group-based user scheduling. We first assume that two-bit feedback is composed of $d$ and instantaneous angle $\varphi$ in the following theorem.
\begin{theorem}\label{the:cdf_twobit_ins}
The CDF of the nonzero square-channel gain for user $i\,{\in}\,\mathcal{S}_{\rm W, \varphi}$ is given as
\begin{align} 
F_{h_i^2|h_i>0}(x) &= \frac{\displaystyle \int_{d^*(x)}^{d_{\rm max}} \Delta \tilde{F}_{\varphi} \left( r,\Theta, \omega \left(x,r,\theta_{\rm th}\right) \right) {\rm d}r}
{\displaystyle \int_{d_{\rm th}}^{d_{\rm max}} \Delta \tilde{F}_{\varphi} \left( r,\Theta, \theta_{\rm th} \right)  {\rm d}r}, \label{eqn:cdf_twobit_ins_i}
\end{align}
where $\Delta \tilde{F}_{\varphi} \left( x,y,z \right) \,{=}\, \Delta F_{\varphi} \left( x,y \right) {-} \Delta F_{\varphi} \left( x,z \right)$, the effective angle is $\omega(x,y,z) \,{=}\, \max \left( 1/2\cos^{{-}1}\!\left( 2\min \left(x\upsilon(y),1\right){-}1\right),z \right)$, and $d^*(x) \,{=}\, u \Big( \sqrt[]{\left( h_c^2 \cos^2\Theta/x \right)^{1/(m{+}2)}{-}\,\ell^2}, d_{\rm th}, d_{\max} \Big)$ with $u(x,y,z) \,{=}\, \min \left( \max \,( x, y ), z \right)$. Similarly, the CDF of the nonzero square-channel gain for user $j\,{\in}\,\mathcal{S}_{\rm S, \varphi}$ is given as
\begin{align} \label{eqn:cdf_twobit_ins_j}
F_{h_j^2|h_j>0}(x) &= 1 {-} \frac{\displaystyle \int_{d_{\rm min}}^{d_{\rm th}} \Delta F_{\varphi} \left( r,\psi \left(x,r,\theta_{\rm th}\right) \right) {\rm d}r}
{\displaystyle \int_{d_{\rm min}}^{d_{\rm th}} \Delta F_{\varphi} \left( r,\theta_{\rm th} \right) {\rm d}r}. 
\end{align}
\end{theorem}
\begin{IEEEproof}
See~\cite{Yapici2018NonVlc} for a complete proof.
\end{IEEEproof}

When the two-bit feedback is computed using the $d$ and the mean angle $\overline{\varphi}$, the desired distribution is given as follows.
\begin{theorem}\label{the:cdf_twobit_mean}
The CDF of the nonzero square-channel for user $i\,{\in}\,\mathcal{S}_{\rm W, \overline{\varphi}}$ is given as
\begin{align} \label{eqn:cdf_twobit_mean_i}
&F_{h_i^2|\overline{h}_i>0}(x) = \frac{\displaystyle A(d^*(x))}{A(d_{\rm th})} + \frac{1}{\Delta\overline{\varphi} A(d_{\rm th})} \times \nonumber \\
& \hspace{0.2in}\int_{d_{\rm th}}^{d^*(x)} \!\!\int_{\mathcal{S}_{\overline{\varphi}}(r)} \! \Big( 1 - \Delta F_{\varphi|\overline{\varphi}} \left( r,\Psi \left(x,r,\Theta\right) \right) \Big) {\rm d}\overline{\varphi}\, {\rm d}r,
\end{align}
where $d^*(x) \,{=}\, u \Big( \sqrt[]{\left( h_c^2/x \right)^{1/(m{+}2)}{-}\,\ell^2} , d_{\rm th}, d_{\max}\Big)$, the effective angle is $\Psi(x,y,z) \,{=}\, \min \left( 1/2\cos^{{-}1}\!\left( 2x\upsilon(y){-}1\right),z \right)$, $\Delta F_{\varphi|\overline{\varphi}} \left( x,y \right)$ is the same as $\Delta F_{\varphi} \left( x,y \right)$ of Theorem~\ref{lem:numof_nzusers} except the additional given mean value $\overline{\varphi}$ for the instantaneous angle ${\varphi}$, and $\mathcal{S}_{\overline{\varphi}}(r) = \left\lbrace \left[ \overline{\varphi}_1(r),\overline{\varphi}_2(r) \right],\left[ \overline{\varphi}_3(r),\overline{\varphi}_4(r) \right] \right\rbrace$ with $\overline{\varphi}_1(r) = \max(\overline{\varphi}_{\min},\alpha({-}\Theta,r)$, $\overline{\varphi}_2(r)= \min(\overline{\varphi}_{\max},\alpha({-}\theta_{\rm th},r))$, $\overline{\varphi}_3(r) = \max(\overline{\varphi}_{\min},\alpha(\theta_{\rm th},r))$, and $\overline{\varphi}_4(r) = \min(\overline{\varphi}_{\max},\alpha(\Theta,r))$, $\alpha(x,r) = \pi {-} \tan^{{-}1}(\ell/r) {+} x$. Moreover, $A(x)$ is defined as
\begin{align}\label{eqn:auxiliary_a}
A(x) &\,{=}\,I(\Theta,x,d_{\max}) \,{-}\, I({-}\Theta,x,d_{\max}) \nonumber\\
&\hspace{0.3in}\,{-}\, I(\theta_{\rm th},x,d_{\max}) \,{+}\, I({-}\theta_{\rm th},x,d_{\max}),
\end{align}
where $I(x,y,z)$ is the indicator function.
Similarly, the CDF of the nonzero square-channel for user $j\,{\in}\,\mathcal{S}_{\rm S, \overline{\varphi}}$ is
\begin{align}
F_{h_j^2|\overline{h}_j>0}(x) &= \frac{B(d^*(x))}{B(d_{\min})} + \frac{1}{\Delta \overline{\varphi}B(d_{\min})} \times\nonumber \\
&\hspace{-0.4in} \int_{d_{\min}}^{d^*(x)} \int_{\overline{\varphi}_2(r)}^{\overline{\varphi}_3(r)} \left( 1{-} \Delta F_{\varphi|\overline{\varphi}} \left( r,\Psi(x,r,\Theta) \right) \right)  {\rm d}\overline{\varphi} \, {\rm d}r,\label{eqn:cdf_twobit_mean_j}
\end{align}
where $B(x) \,{=}\, I(\theta_{\rm th},x,d_{\rm th}) \,{-}\, I({-}\theta_{\rm th},x,d_{\rm th})$.
\end{theorem}
\begin{IEEEproof}
See~\cite{Yapici2018NonVlc} for a complete proof.
\end{IEEEproof}

As before, the desired outage probabilities of group-based user-scheduling NOMA given in \eqref{eqn:outpr_i_3} and \eqref{eqn:outpr_j_3} can then be computed using the respective nonzero square-channel CDFs given in Theorem~\ref{the:cdf_twobit_ins} and Theorem~\ref{the:cdf_twobit_mean}.

\vspace{-0.15in}
\section{Numerical Results}\label{sec:results}
In this section, we present numerical results for the the sum-rate performance of the NOMA strategies and feedback schemes considered in Section~\ref{sec:noma}. In this regard, we assume a total of $K\,{=}\,20$ users, each of which has $d_{\min}\,{=}\,0\,\text{m}$, $d_{\max}\,{=}\,10\,\text{m}$, $\overline{\varphi}_\text{min}\,{=}\,\Delta\varphi$, $\overline{\varphi}_\text{max}\,{=}\,180-\Delta\varphi$, so that the instantaneous vertical angle spans $[0^\circ,180^\circ]$ irrespective of the particular $\Delta\varphi$ value. We also assume that the LED is vertically off the horizontal plane by $\ell\,{=}\,2\,\text{m}$ with $\Phi_{\rm HPBW}\,{=}\,60^\circ$, and the photodetector has $A_{\rm e}\,{=}\,1\,\text{cm}^2$ with a FOV of $100^\circ$ (i.e., $\Theta\,{=}\,50^\circ$). We choose the power allocation coefficients to be $\beta_i\,{=}\,63/64$ and $\beta_j\,{=}\,1/64$ along with the respective target rates of $\overline{R}_i\,{=}\,2\,\text{bit/s/Hz}$ and $\overline{R}_j\,{=}\,10\,\text{bit/s/Hz}$. 

\begin{figure}[!t]
\centering
\includegraphics[width=0.5\textwidth]{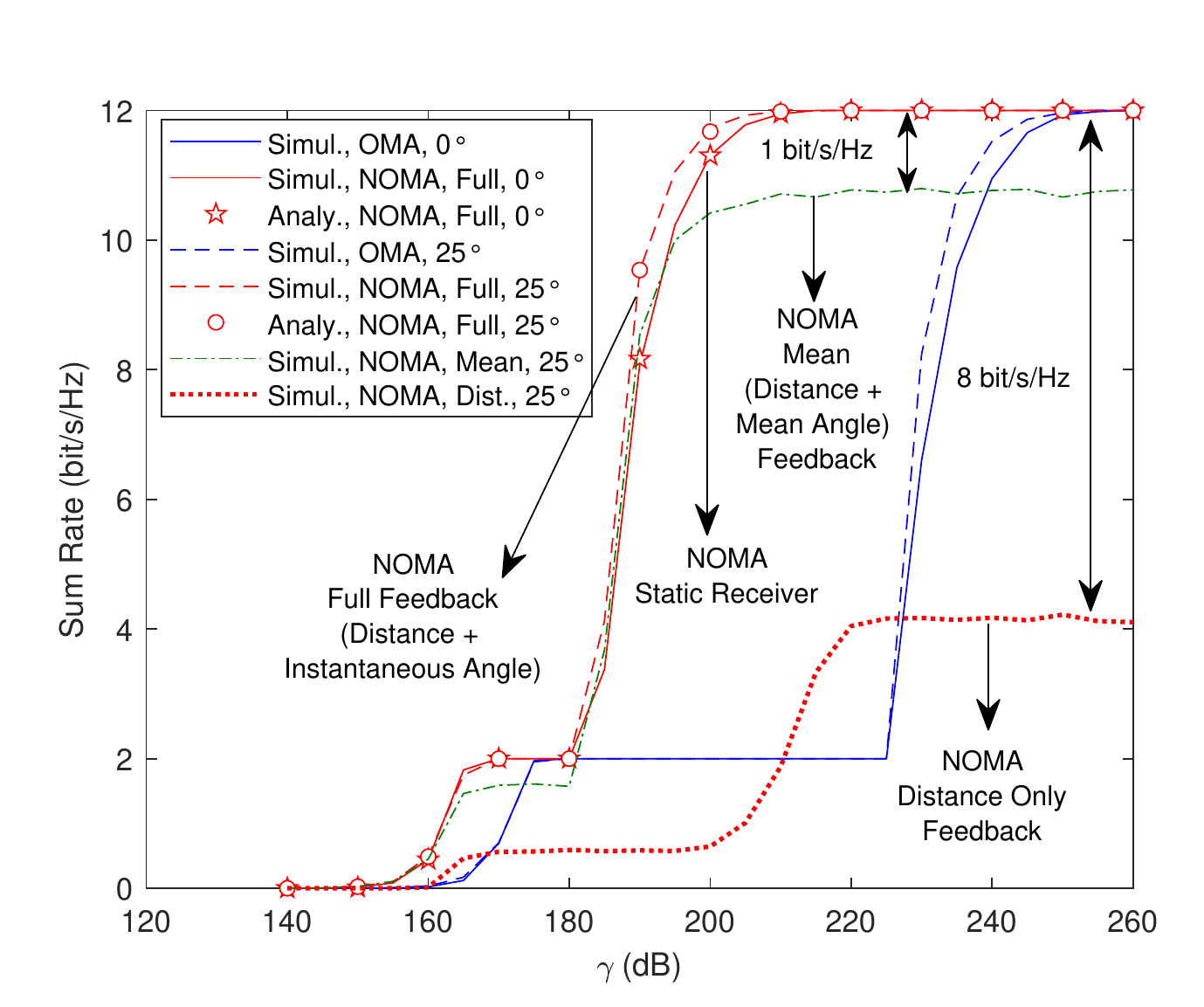}
\caption{OMA and NOMA sum rates against transmit SNR ($\gamma$) with individual user scheduling strategy, where $\Delta\varphi\,{=}\,\{0^\circ,25^\circ\}$ and $\text{FOV}\,{=}\,100^\circ$.}
\label{fig:nogroup_sumrate}\vspace{-0.0in}
\end{figure}

In Fig.~\ref{fig:nogroup_sumrate}, we depict the sum rates of OMA and NOMA with the individual user scheduling for $i\,{=}\,1$, $j\,{=}\,10$, $\Delta\varphi\,{=}\,\{0^\circ,25^\circ\}$. Note that while $\Delta\varphi\,{=}\,0^\circ$ corresponds to ``\textit{static}'' receiver orientation in the vertical domain, $\Delta\varphi\,{=}\,25^\circ$ represents ``\textit{dynamic}'' receiver orientation with a large variation (i.e., orientation spans as large as $50^\circ$ in the vertical domain over time). We observe that NOMA outperforms OMA in terms sum rates, and that analytical results nicely follow the simulation data in all the cases. We also observe that the performance of the mean vertical angle based limited feedback is very close to that of the full CSI based feedback (even for a large $\Delta\varphi$ value of $25^\circ$). On the other hand, the distance only feedback cannot capture the FOV status correctly when $\Delta\varphi$ gets larger, and the steady-state sum-rate loss is $8$~\mbox{bit/s/Hz}.

\begin{figure}[!t]
\centering
\includegraphics[width=0.5\textwidth]{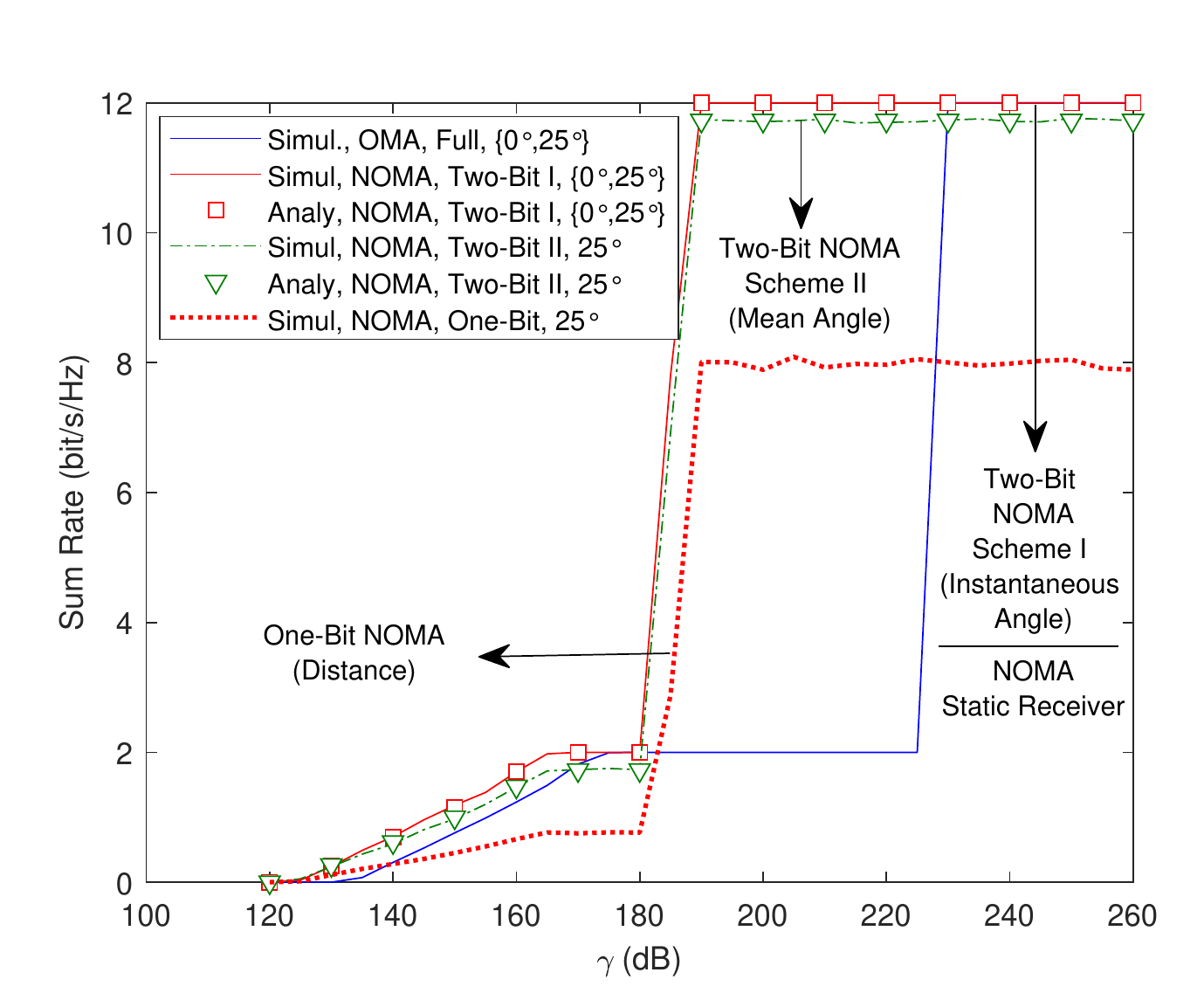}
\label{fig:twobit_sumrate_fov100}
\caption{OMA and NOMA sum rates against transmit SNR ($\gamma$) with group-based user scheduling, $\Delta\varphi\,{=}\,\{0^\circ,25^\circ\}$, $\text{FOV}\,{=}\,100^\circ$, $d_{\rm th} {=}\, 1\,\text{m}$, $\theta_{\rm th} {=}\,5^\circ$.}
\label{fig:twobit_sumrate}\vspace{-0.0in}
\end{figure}

We now consider the sum-rate performance of NOMA with the group-based user scheduling. We assume that $d_{\rm th} \,{=}\, d_{\min} \,{+}\, c_{d_{\rm th}} \left( d_{\max} \,{-}\, d_{\min} \right)$ and $\theta_{\rm th} \,{=}\, c_{\theta_{\rm th}} \Theta$, where $c_{d_{\rm th}} {\in}\, [0,1]$ and $c_{\theta_{\rm th}} {\in}\, [0,1]$ are the threshold coefficients to determine the values of $d_{\rm th}$ and $\theta_{\rm th}$, respectively. In Fig.~\ref{fig:twobit_sumrate}, we plot the respective sum-rate results of OMA and NOMA with varying transmit SNR, where $c_{d_{\rm th}} \,{=}\, 0.1$, $c_{\theta_{\rm th}} \,{=}\, 0.1$, and $\Delta\varphi\,{=}\,\{0^\circ,25^\circ\}$. As before, the analytical results nicely matches the experimental data, and NOMA achieves better sum-rate performance than OMA. We observe that the sum-rate performance for NOMA with two-bit feedback of $d$ and $\varphi$ (referred to as Scheme I) remains the same as $\Delta\varphi$ increases from $0^\circ$ to $25^\circ$. Two-bit feedback based NOMA is therefore very robust to the random receiver orientation. When $\overline{\varphi}$ is employed instead of $\varphi$ in the feedback computation (referred to as Scheme II), the degradation in NOMA sum rates is less than $0.3$~\mbox{bit/s/Hz} at the steady state. This result underscores the power of the practical feedback scheme for two-bit feedback based NOMA involving mean vertical angle (instead of using its instantaneous value). 

\begin{figure}[!t]
\centering
\includegraphics[width=0.5\textwidth]{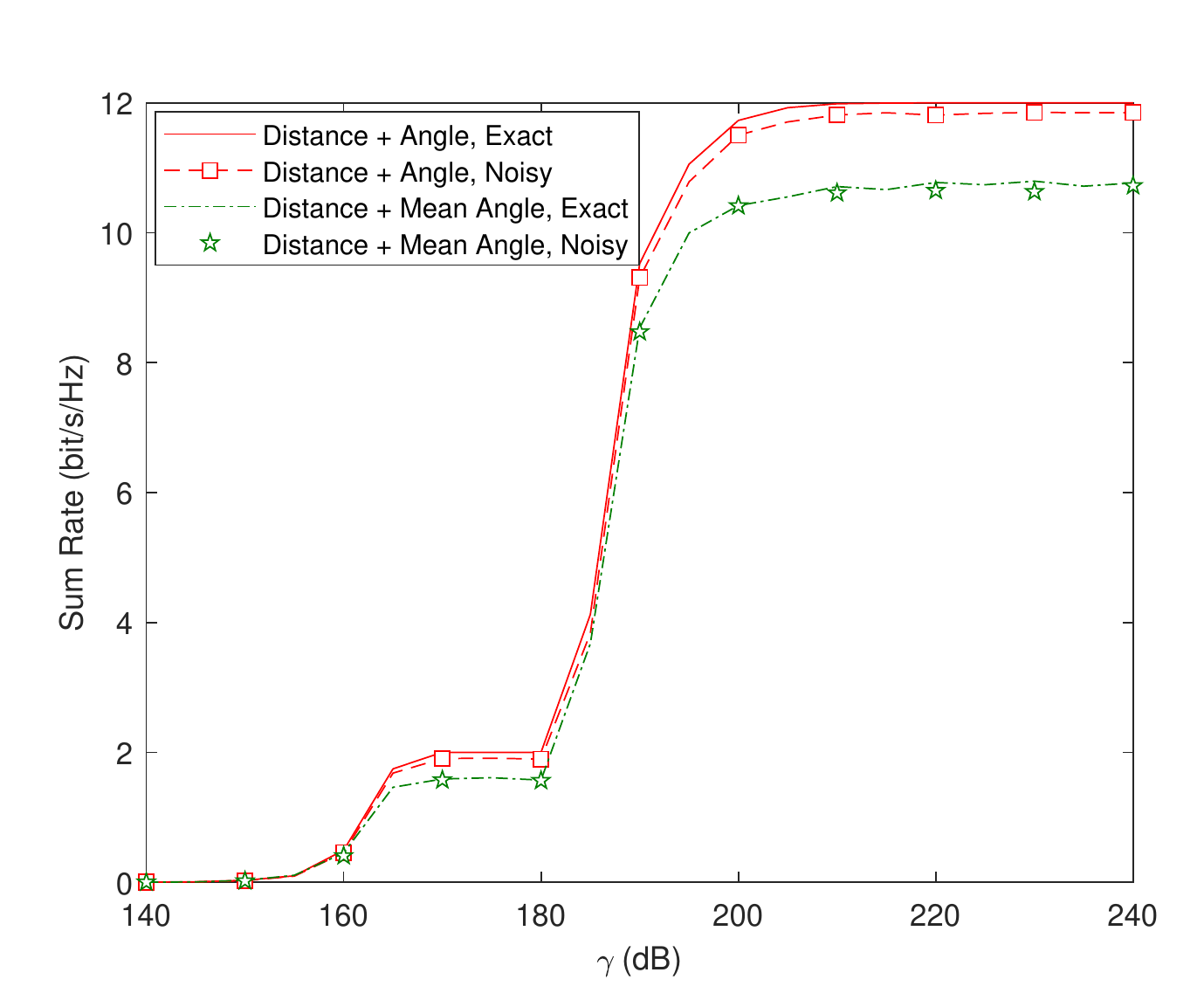}
\caption{Simulation results for NOMA sum rates against transmit SNR ($\gamma$) with individual user scheduling strategy, where distance and angle error for noisy case is $0.1\,\text{m}$ and $5^\circ$, respectively,    $\Delta\varphi\,{=}\,25^\circ$ and $\text{FOV}\,{=}\,100^\circ$. }
\label{fig:nogroup_sumrate_noisy}
\vspace{-0.2in}
\end{figure}

We finally consider the impact of \textit{noisy} horizontal distance and vertical angle information on the NOMA sum rates in a \textit{dynamic} scenario with $\Delta\varphi\,{=}\,25^\circ$. To this end, we consider noisy estimates as $\hat{d}_k \,{=}\, d_k \,{+}\, \epsilon_d$, $\hat{\varphi}_k \,{=}\, \varphi_k \,{+}\, \epsilon_\varphi$, and $\hat{\overline{\varphi}}_k \,{=}\, \overline{\varphi}_k \,{+}\, \epsilon_\varphi$, where $\epsilon_d$ and $\epsilon_\varphi$ stand for estimation error, and are assumed to be complex Gaussian with zero-mean and variance $\sigma_d^2$ and $\sigma_\varphi^2$, respectively. We assume that $\sigma_d \,{=}\, 0.05$ and $\sigma_\varphi \,{=}\, 2.5$, which correspond to  $0.1\,\text{m}$ error in distance and $5^\circ$ error in vertical angle~\cite{Lau2018SimPos}. In Fig.~\ref{fig:nogroup_sumrate_noisy}, we depict the simulation results for individual user-scheduling NOMA with noisy distance and angle information assuming the same setting of Fig.~\ref{fig:nogroup_sumrate}. We observe that performance of the mean vertical angle based limited-feedback scheme does not change while that of the instantaneous angle based scheme exhibits a marginal degradation.

\vspace{0.0in}
\section{Conclusion}\label{sec:conclusion}
We investigated a downlink multiuser VLC scenario involving mobile users with random vertical orientation. In order to increase the spectral efficiency, the NOMA transmission is employed with various user scheduling techniques and feedback mechanisms. The outage probability and sum-rate expressions are derived analytically, where the respective numerical results show a very good match with the simulation data. We observe that the practical feedback scheme with the mean vertical angle achieves a near-optimal sum-rate performance. In addition, the two-bit feedback involving both the distance and the angle information significantly outperforms the conventional one-bit feedback with the distance information only.

\bibliographystyle{IEEEtran}
\bibliography{IEEEabrv,papers}

\end{document}